\documentclass[12pt,a4paper]{article}
\pdfoutput=1
\usepackage{macros}
\usepackage{amssymb}

\newcommand{\nn}{\nonumber}

\preprint{}
\title{Bethe/Gauge Correspondence for $A_{N}$ Spin Chains with Integrable Boundaries }

\author[a]{Ziwei Wang}
\author[a,b]{Rui-Dong Zhu}
\affiliation[a]{School of Physical Science and Technology, Soochow University, 333 Ganjiang Road, 215006 Suzhou, China
}
\affiliation[b]{Institute for Advanced Study, Soochow University, 333 Ganjiang Road, 215006 Suzhou, China}

\abstract{We continue the survey initiated in \cite{Kimura:2020bed} to explore the Bethe/Gauge correspondence between supersymmetric SO/Sp gauge theories in 2d/3d/4d and open spin chain with integrable boundaries. We collect the known Bethe ansatz equations of different types of spin chains with general boundaries that have been analyzed in the literature, and compare them with the vacua equations of the quiver gauge theories. It seems that not all the vacua equations of quiver gauge theory with BCD-type gauge groups can be realized as some known Bethe ansatz equations of integrable spin chain models. }

\begin{document}

\allowdisplaybreaks

\maketitle

\section{Introduction}

The connection between supersymmetric gauge theories and integrable models has been known and intensively studied ever since the classical work of Seiberg and Witten \cite{Seiberg:1994rs,Seiberg:1994aj}. It is, however, after the work of Nekrasov and Shatashvili \cite{Nekrasov:2009uh,Nekrasov:2009ui}, people started to believe that quantum integrability exists more widely behind the mathematical structure of supersymmetric gauge theories. In Nekrasov-Shatashvili's work, a new kind of dualities were proposed between 2d $\cN=(2,2)$ U($k$) gauge theories and Heisenberg XXX spin chains, and now such a duality is often referred to as the Bethe/Gauge correspondence. The correspondence has then been explored and tested from many aspects, e.g. the wavefunction \cite{Bullimore:2017lwu,Foda:2019klg}, the R-matrix \cite{Bullimore:2017lwu,Gu:2022ugf}, rational Q-systems \cite{Gu:2022dac} and the number of solutions \cite{Shu:2022vpk} etc., and it is expected to be naturally promoted to the correspondences between 3d $\cN=2$ (resp. 4d $\cN=1$) gauge theories and XXZ (resp. XYZ) integrable spin chains. The vortex partition function of 2d $\cN=(2,2)$ U($k$) gauge theories, which gives the effective twisted potential used in the establishment of the Bethe/Gauge correspondence, can be derived from the Nekrasov partition function at the root of the Higgs branch of 4d $\cN=2$ gauge theory \cite{Fujimori:2015zaa}. Associated to the Nekrasov partition function, moreover, a quantum algebra called the affine Yangian of $\widehat{\mathfrak{gl}}_1$ \cite{Schiffmann:2012tbu} (or the quantum toroidal algebra of $\widehat{\mathfrak{gl}}_1$ \cite{Awata:2011ce}) is found in 4d (or resp. 5d) instanton moduli space and an R-matrix usually called Maulik-Okounkov's R-matrix is attached \cite{Maulik:2012wi}. The algebraic structure and the R-matrix, which allows one to build an integrable system based on Fock spaces, implies the integrable nature of the supersymmetric gauge theories with U($k$) gauge groups on the full $\Omega$-background (a good review on the quantum toroidal algebra can be found in \cite{Matsuo:2023lky}). When it comes to other types of gauge groups, the situation becomes much less clear. One of the main reasons is the complication appearing in the evaluation of the Nekrasov partition function of gauge theories beyond the A-type. Despite the ADHM construction of instantons available in all classical gauge groups \cite{Atiyah:1978ri,Nekrasov:2004vw,Kim:2012gu}, the classification of Jeffrey-Kirwan poles in the localization calculation of the Nekrasov partition function turned out to be messy \cite{Hollands:2010xa,Nakamura:2014nha,Nakamura:2015zsa} and it is hard to find a closed-form analytic expression for the instanton partition function. People have so far only found in the self-dual $\Omega$-background some analytic expressions for the instanton partition functions of BCD-type gauge theories \cite{Kim:2017jqn,Hayashi:2020hhb,Nawata:2021dlk} based on the generalization of the topological vertex formalism, and a very preliminary candidate for the underlying quantum algebra in the self-dual limit was proposed in \cite{Nawata:2023wnk}. It is thus still very far from the explicit confirmation or denying of the quantum integrability in BCD-type gauge theories. 

In this article, we take one step back to focus on the proposal of the Bethe/Gauge correspondence briefly made in the original work of \cite{Nekrasov:2009uh,Nekrasov:2009ui} that gauge theories with BCD-type gauge groups might be dual to spin chains with open boundaries. The details of this proposal was given in \cite{Kimura:2020bed} and we wish to further extend and examine such Bethe/Gauge correspondences as thoroughly as possible in the cases of quiver gauge theories and 4d $\cN=1$ theories with BCD-type gauge groups. Note that since the Bethe ansatz equation of Maulik-Okounkov's R-matrix, which was derived in \cite{Feigin:2015raa,Litvinov:2020zeq}, reproduces that of the Heisenberg spin chain in a special limit, it is believed that the establishment of the Bethe/Gauge correspondence is a necessary condition to have the quantum integrability in the more general parameter region of gauge theories beyond the A-type. A supportive hint for this was provided by the open integrable system constructed in \cite{Litvinov:2021phc} with the boundary operator defined associated to Maulik-Okounkov's R-matrix, and again the Bethe ansatz equation obtained there reduces to that of Heisenberg spin chains with open boundaries. 

The article is organized as the following. In section \ref{s:closed}, we provide a brief review on the Bethe/Gauge correspondence for spin chains with (twisted) periodic boundary condition, and further work out some details of the correspondence between 4d gauge theories and XYZ spin chains missing in the literature. In section \ref{s:bethe-open}, we present the Bethe ansatz equations of various kinds of open spin chains to discuss the candidate quiver gauge theories that can potentially be realized, including open spin chains with off-diagonal boundaries, $A_N$ XXZ spin chains with diagonal boundaries, $A_N$ XXX spin chains with general boundaries and XYZ open spin chains with general boundaries, and then we summarize the results with some further discussions presented in section \ref{s:conclusion}.

\section{Rank-1 Bethe/Gauge correspondence to closed spin chains}\label{s:closed}

In this section, we mainly review the duality first proposed in \cite{Nekrasov:2009uh,Nekrasov:2009ui} between the vacua of 2d/3d/4d gauge theory and Bethe states of XXX/XXZ/XYZ spin chain. We focus on the construction of such dualities in the rank-1 closed spin chains in this section with some details re-investigated, and we will consider the further generalization to open chains with focus on higher-rank cases in the next section.

\subsection{3d and 2d version corresponding to closed chains}\label{s:BG-KZ}

The 3d version of the Bethe/Gauge correspondence proposed originally in \cite{Nekrasov:2009uh,Nekrasov:2009ui} connects 3d $\cN=2$ U($k$) gauge theories of some particular type to Heisenberg XXZ spin chain with twisted periodic boundary condition. 

At the most primitive level, we aim to match the Bethe ansatz equation (BAE) of the spin chain with the vacua equation of the gauge theory. On the one hand, we consider the XXZ Heisenberg spin chain, whose Hamiltonian is of the form, 
\begin{equation}
    {\cal H}=-\frac{1}{2}\sum_{i=1}^L\lt(J_x^{(i)}J_x^{(i+1)}+J_y^{(i)}J_y^{(i+1)}+\cosh(\eta)J_z^{(i)}J_z^{(i+1)}\rt),
\end{equation}
with twisted periodic boundary condition, 
\begin{equation}
    J_z^{(L+1)}=J_z^{(1)},\quad J^{(L+1)\pm}=e^{\pm i\theta}J^{(1)\pm},\label{bc}
\end{equation}
for a twisted angle $\theta$. The corresponding transfer matrix can be constructed as 
\begin{align}
    t(u)&:={\rm Tr}_0\lt({\rm Ph}_2(\theta)L_{0L}(u-\theta_L)L_{0(L-1)}(u-\theta_{L-1})\dots L_{02}(u-\theta_2)L_{01}(u-\theta_1)\rt)\cr
    &=:{\rm Tr}_0\lt[{\rm Ph}_2(\theta)\lt(\begin{array}{cc}
    {\cal A}^{(L)}u) & {\cal B}^{(L)}(u)\\
    {\cal C}^{(L)}(u) & {\cal D}^{(L)}(u)\\
    \end{array}
    \rt)\rt]=e^{i\theta/2}{\cal A}^{(L)}(u)+e^{-i\theta/2}{\cal D}^{(L)}(u),\label{def-transfer}
\end{align}
where $L_{ij}$ is the Lax matrix satisfying the RLL relation 
\begin{equation}
    R_{12}(u-v)L_{13}(u)L_{23}(v)=L_{23}(v)L_{13}(u)R_{12}(u-v),\label{RLL}
\end{equation}
associated to a given R-matrix $R(u)$ and $\{\theta_{i}\}_{i=1}^L$ are the inhomogeneous parameters. By finding the eigenvalue of the transfer matrix, one naturally diagonalizes all the conserved charges $\mathbb{Q}_i$'s, given as the expansion coefficients of $t(\lambda)$, 
\begin{equation}
    t(u)=\sum_{i=0}^\infty \mathbb{Q}_iu^i.
\end{equation}
Among them, the Hamiltonian can be found in our convention as 
\begin{equation}
    {\cal H}=-i\left.\frac{{\rm d}}{{\rm d}u}\log t(u)\right|_{u=\frac{i}{2}}.
\end{equation}
The eigenstates to the transfer matrix take the form, 
\begin{equation}
    \ket{\Psi(L,m)}={\cal B}^{(L)}(u_1){\cal B}^{(L)}(u_2)\dots{\cal B}^{(L)}(u_m)\ket{\Omega},\label{ABA-A1}
\end{equation}
where the ground state $\ket{\Omega}$ is annihilated by ${\cal C}^{(L)}(u)$, i.e. ${\cal C}^{(L)}(u)\ket{\Omega}=0$. Only when the set of parameters $\{u_i\}_{i=1}^m$, usually called the Bethe roots, satisfy the following Bethe ansatz equation (BAE), 
\begin{equation}
    \prod_{a=1}^L\frac{\sinh(u_i+\eta/2+\eta s_a-\theta_a)}{\sinh(u_i+\eta/2-\eta s_a-\theta_a)}=e^{i\theta}\prod_{j\neq i}^m\frac{\sinh(u_i-u_j+\eta)}{\sinh(u_i-u_j-\eta)},\label{BAE-p}
\end{equation}
the state presented in \eqref{ABA-A1} becomes a true eigenstate of the system. 

On the other hand, one derives the vacua equation of the gauge theory from the following effective twisted potential, 
\begin{align}
    &W^{3d}_{eff}(\sigma,m)=-\frac{1}{\beta_2}\sum_\alpha Li_2(e^{i\alpha\cdot \sigma})+\frac{1}{4\beta_2}\sum_\alpha (\alpha\cdot\sigma)^2\cr
&+\frac{1}{\beta_2}\sum_w \sum_{a=1}^{N_f}Li_2(e^{-iw\cdot \sigma-im_a-i\beta_2\tilde{c}})-\frac{1}{4\beta_2}\sum_w \sum_{a=1}^{N_f}(w\cdot\sigma+m_a+\beta_2\tilde{c})^2.\label{W-3d}
\end{align}
Here we followed the notations used in \cite{Yoshida:2014ssa,Kimura:2020bed} in the computation of the $D^2\times S^1$ partition function of 3d $\cN=2$ gauge theories. $\alpha$ runs over all the roots in the gauge group, $w$ stands for the weights appearing in the representation of each chiral multiplet. 
When the gauge group contains a U(1) subgroup, one can further introduce an FI-term, 
\begin{equation}
    W_{\rm FI}(\sigma)=2\pi i\ell \zeta{\rm tr}\sigma.
\end{equation}
added to the effective twisted potential \eqref{W-3d}. 
The vacua equation is then derived from 
\begin{equation}
    \exp\lt(\beta_2 i\frac{\partial}{\partial\sigma}W^{3d}_{eff}(\sigma,m)\rt)=1,\label{vacua-eq}
\end{equation}
and in the simplest case of U($k$) gauge theory with one adjoint matter (with mass $m_{adj}$), $N_f$ fundamental matters (with mass parameters $\{m_a\}_{a=1}^{N_f}$) and $N'_f$ anti-fundamental matters (with masses $\{\bar{m}_a\}_{a=1}^{N_f}$), the vacua equation is given by
\begin{equation}
    (-1)^{N'_f}e^{-2\pi \beta_2\ell\zeta}\prod_{j\neq i}^k\frac{[\sigma_i-\sigma_j+m_{adj}+\beta_2\tilde{c}]}{[\sigma_i-\sigma_j-m_{adj}-\beta_2\tilde{c}]}\frac{\prod_{a=1}^{N'_f}[\sigma_i-\bar{m}_a-\beta_2\tilde{c}]}{\prod_{a=1}^{N_f}[\sigma_i+m_a+\beta_2\tilde{c}]}=1,\label{vacua-U}
\end{equation}
where $[x]:=2\sin\lt(\frac{x}{2}\rt)$. For simplicity, one can shift the mass parameters to set the fugacity $\tilde{c}$ to zero in the below. According to the Bethe/Gauge correspondence proposed in \cite{Nekrasov:2009uh,Nekrasov:2009ui}, one can write down a map to match two equations of \eqref{BAE-p} and \eqref{vacua-U}. The dictionary between the closed XXZ quantum spin chain and the 3d U($k$) gauge theory with equal number $N_f$ of fundamental and anti-fundamental matters is given by 
\begin{align}
&m\leftrightarrow k,\quad L\leftrightarrow N_f,\quad u_j\leftrightarrow i\sigma_j/2,\quad  \eta\leftrightarrow im_{adj}/2,\quad i\theta \leftrightarrow \pi i N_f-2\pi \beta_2\ell\zeta,\label{U-map}
\end{align}
and the mass parameters need to be specified as
\begin{align}
(\eta/2+\eta s_a-\theta_a)\leftrightarrow i m_a/2,\quad (\eta/2-\eta s_a-\theta_a)\leftrightarrow -i\bar{m}_a/2.
\end{align}

One can rescale the Bethe roots $u_i$ and the parameter $\eta$ as 
\begin{equation}
    u_i=\tilde{u}_i\delta,\quad \eta=i\delta,
\end{equation}
and take $\delta\rightarrow 0$ to obtain a corresponding XXX spin chain with the same boundary condition from the XXZ spin chain. This simply modifies the trigonometric function $\sinh(u)$ in the BAE to its rational version $\tilde{u}$. Accordingly, by taking the radius $\beta \ell$ of $S^1$ of the 3d gauge theory to zero, one recovers a 2d theory on the gauge side. The gauge theory parameters are rescaled to $\sigma=\beta\ell \tilde{\sigma}$, $m=\beta\ell\tilde{m}$, and one can effectively replace $[x]\rightarrow \tilde{x}$ in the vacua equation. The Bethe/Gauge correspondence thus holds completely in parallel in 3d and 2d for closed spin chains.

\subsection{4d version corresponding to periodic XYZ spin chains}

It was also proposed in \cite{Nekrasov:2009uh,Nekrasov:2009ui} that the Bethe/Gauge correspondence can be generalized to that between 4d gauge theories and the XYZ spin chain. Interestingly the XYZ spin chain is technically rather different from the XXZ and XXX spin chain, as it does not possess a U(1) symmetry usually associated to the number of magnons $m$. A review with some more details on the correspondence to periodic XYZ spin chains is provided in this section. Due to the limited results available in the literature to spin chains with spin larger than $\frac{1}{2}$, we only focus on the case that the spin at each site of the spin chain is taken to $s=\frac{1}{2}$ in the remaining of this article unless declared otherwise. We note, however, that it is expected that one can replace the combination $-\theta_i\pm \eta/2$ in the BAE by $-\theta_i\pm s_i\eta$ for $\theta_i$ and $s_i$ respectively standing for the inhomogeneous parameter and the spin at each site of the spin chain to obtain the BAE in the most general case.  

\subsubsection{Vacua equation from 4d}

The vacua equation of 4d $\cN=1$ gauge theory can be derived from its effective potential deduced from the index on $D^2\times T^2$ available in \cite{Longhi:2019hdh} (see Appendix \ref{a:4d-part}), 
\begin{align}
    W^{4d}_{eff}(\Phi)=-\sum_\alpha g(\alpha\cdot\Phi;p)-\frac{\pi^2}{3\beta_2\varsigma}\sum_{\alpha}(\alpha\cdot\Phi)^3+\frac{\pi^2(1+\varsigma)}{2\beta_2\varsigma}\sum_\alpha(\alpha\cdot \Phi)^2\cr
+\sum_w\sum_f g(w\cdot\Phi+2\pi im_f;p)+\frac{\pi^2}{3\beta_2\varsigma}\sum_{w}\sum_f(w\cdot\Phi+m_f)^3\cr
-\frac{\pi^2(1+\varsigma)}{2\beta_2\varsigma}\sum_w\sum_f(w\cdot \Phi+m_f)^2,\label{4d-potential}
\end{align}
where $p:=e^{2\pi i\varsigma}$ with $\varsigma$ the modulus parameter of $T^2$ and 
\begin{equation}
    g(x;p):=\frac{1}{2\beta_2}\sum_{k=1}^\infty \frac{e^{2\pi ik x}}{k^2(1-p^k)}-\frac{1}{2\beta_2}\sum_{k=1}^\infty \frac{p^ke^{-2\pi ik x}}{k^2(1-p^k)}.
\end{equation}
$\alpha$ in the effective potential again runs over all the roots in the gauge group and $w$ stands for the weights appearing in the representation of each chiral multiplet. In the case of U($k$) theory with $N_f$ fundamental and anti-fundamental multiplets, the vacua condition in 4d given by 
\ba
\exp\lt(2\beta_2 \frac{\partial}{\partial(2\pi i\Phi_j)}W^{4d}_{eff}(\{\Phi_j\},\{m_f\})\rt)=1,
\ea
is found to be 
\begin{align}
    \prod_{j\neq i} \frac{\theta(e^{2\pi i(\Phi_i-\Phi_j)};p)}{\theta(e^{2\pi i(\Phi_j-\Phi_i)};p)}\frac{\theta(e^{2\pi i(\Phi_j-\Phi_i)+2\pi im_{adj}};p)}{\theta(e^{2\pi i(\Phi_i-\Phi_j)+2\pi im_{adj}};p)}e^{-\frac{4\pi i}{\varsigma}(\Phi_i-\Phi_j)m_{adj}}\nn\\
=e^{\frac{\pi i}{\varsigma}(\sum_fm_f^2-\sum_{\bar{f}}\bar{m}^2_{\bar{f}})-\frac{\pi i(1+\varsigma)}{\varsigma}(\sum_f m_f-\sum_{\bar{f}}\bar{m}_{\bar{f}})}e^{-\frac{2\pi i(1+\varsigma)}{\varsigma}N_f\Phi_i}\cr
\times\frac{\prod_f\theta(e^{2\pi i(\Phi_i+m_f)};p)e^{\frac{2\pi i}{\varsigma}\Phi_i m_f}}{\prod_{\bar{f}}\theta(e^{2\pi i(-\Phi_i+\bar{m}_{\bar{f}})};p)e^{-\frac{2\pi i}{\varsigma}\Phi_i\bar{m}_{\bar{f}}}},
\end{align}
where we used
\begin{equation}
    2\beta_2\frac{\partial}{\partial(2\pi iz)}g(z;p)=-\sum_{i=0}^\infty \log(1-e^{2\pi iz}p^i)-\sum_{i=0}^\infty \log(1-e^{-2\pi iz}p^{i+1}),
\end{equation}
and therefore 
\begin{equation}
    \exp\lt(2\beta_2\frac{\partial}{\partial(2\pi iz)}g(z;p)\rt)=\frac{1}{\theta_0(z,\tau)},
\end{equation}
with the convention $\theta_0(z,p):=(x,p)(px^{-1},p)$, and $x:=\exp(2\pi iz)$ in the $\theta_0$-function. 
It can be further rewritten in terms of Jacobi's $\vartheta$-function, often also denoted as $\sigma(z)\equiv\vartheta_{1,1}(z,p^{\frac{1}{2}})$ depending on the context, 
\ba
\sigma(u):=-2p^{\frac{1}{8}}\sin(\pi u)\prod_{m=1}^\infty (1-p^{m})(1-e^{2\pi iu}p^{m})(1-e^{-2\pi iu}p^{m}),
\ea
and we have 
\begin{equation}
    \sigma(u)=-ip^{\frac{1}{8}}(p,p)u^{-\frac{1}{2}}\theta_0(u,p),\quad (p,p):=\prod_{m=1}^\infty (1-p^{m}).
\end{equation}
The vacua equation then becomes 
\begin{align}
    &\prod_{j\neq i} \frac{\sigma(\Phi_j-\Phi_i+m_{adj})}{\sigma(\Phi_i-\Phi_j+m_{adj})}\frac{\prod_{\bar{f}}\sigma(-\Phi_i+\bar{m}_{\bar{f}})}{\prod_f\sigma(\Phi_i+m_f)}\cr
&=(-1)^{k-1}e^{\frac{\pi i}{\varsigma}(\sum_fm_f^2-\sum_{\bar{f}}\bar{m}^2_{\bar{f}})-\frac{\pi i(1+\varsigma)}{\varsigma}(\sum_f m_f-\sum_{\bar{f}}\bar{m}_{\bar{f}})}e^{E(\Phi_i;m_f,\bar{m}_{\bar{f}},m_{adj})},\label{4d-vacua}
\end{align}
where we defined 
\begin{align}
    e^{E(\Phi_i;m_f,\bar{m}_{\bar{f}},m_{adj})}=e^{-\frac{2\pi i}{\varsigma}N_f\Phi_i}\prod_f e^{\frac{2\pi i}{\varsigma}\Phi_i m_f}\prod_{\bar{f}}e^{\frac{2\pi i}{\varsigma}\Phi_i\bar{m}_{\bar{f}}}\prod_{j\neq i}e^{\frac{4\pi i}{\varsigma}(\Phi_i-\Phi_j)m_{adj}}\cr
=e^{-\frac{2\pi i}{\varsigma}N_f\Phi_i} e^{\frac{2\pi i}{\varsigma}\Phi_i \lt(\sum_f m_f+\sum_{\bar{f}}\bar{m}_{\bar{f}}\rt)}e^{\frac{4\pi i}{\varsigma} \lt(k\Phi_im_{adj}-\Phi_{U(1)}\rt)}.\label{def-Ephi}
\end{align}
When the gauge group contains a U(1) subgroup, it is also possible to turn on an FI parameter $\zeta$ as in the 3d case and we can add a constant term in the above function of $E$ to $E(\Phi_i;m_f,\bar{m}_{\bar{f}},m_{adj};\zeta)$. 
There is a modularity condition discussed in \cite{Nekrasov:2009uh} stating that 
\begin{equation}
km_{adj}+(\sum_f m_f+\sum_{\bar{f}}\bar{m}_{\bar{f}})=0,\label{modularity} 
\end{equation}
which constrains the field contents of the gauge theory. 

\subsubsection{BAE of XYZ spin chain and the correspondence}

Unlike the rather parallel computation done in 4d, some drastic change occurs when we go to the XYZ spin chain. In particular, the usual U(1) symmetry associated to the magnon number is missing and the construction of the ground state also becomes involved (more details can be found e.g. in \cite{Baxter:1982zz}). In general, the TQ-relation contains three parts, and the BAE is thus also modified to a nontraditional form \cite{Cao:2013gug}, which can be found from the off-diagonal Bethe ansatz approach developed in \cite{Cao:2013nza}. However in some special cases, it is still possible to put the third term in the TQ-relation to zero and obtain a BAE of the traditional type. In this article, we only focus on such cases that can be easily compared to the vacua equation of the gauge theory. 

Another different point is that the twisted boundary condition realized by inserting ${\rm Ph}_2(\theta)$ as in \eqref{def-transfer} is not allowed in the XYZ model as ${\rm Ph}_2(\theta)$ is no longer a solution to the RLL relation \eqref{RLL} when we use the R-matrix, 
\begin{equation}
    {\bf R}(u)=\lt(\begin{array}{cccc}
\alpha(u) & & & \delta(u)\\
& \beta(u) & \gamma(u) & \\
& \gamma(u) & \beta(u) & \\
\delta(u) & & & \alpha(u)\\
\end{array}\rt),\label{R-XYZ}
\end{equation}
where
\begin{align}
    \alpha(u)=\frac{\theta_{0,1/2}(u,2\tau)\theta_{1/2,1/2}(u+\eta,2\tau)}{\theta_{0,1/2}(0,2\tau)\theta_{1/2,1/2}(\eta,2\tau)},\quad \beta(u)=\frac{\theta_{1/2,1/2}(u,2\tau)\theta_{0,1/2}(u+\eta,2\tau)}{\theta_{0,1/2}(0,2\tau)\theta_{1/2,1/2}(\eta,2\tau)},\\
\gamma(u)=\frac{\theta_{0,1/2}(u,2\tau)\theta_{0,1/2}(u+\eta,2\tau)}{\theta_{0,1/2}(0,2\tau)\theta_{0,1/2}(\eta,2\tau)},\quad \delta(u)=\frac{\theta_{1/2,1/2}(u,2\tau)\theta_{1/2,1/2}(u+\eta,2\tau)}{\theta_{0,1/2}(0,2\tau)\theta_{0,1/2}(\eta,2\tau)},
\end{align}
with 
\ba
\theta_{a_1,a_2}(u,\tau)=\sum_{m=-\infty}^\infty \exp\lt(i\pi \lt((m+a_1)^2\tau+2(m+a_1)(u+a_2)\rt)\rt).
\label{theta_fn}
\ea
In the case of the periodic XYZ spin chain, the monodromy matrix is again parameterized by $L$ inhomogeneous parameters $\theta_{i=1,\dots,L}$. 
\begin{equation}
    {\bf T}_0(u)={\bf R}_{0L}(u-\theta_L)\dots {\bf R}_{01}(u-\theta_1).
\end{equation}
Let us first summarize the results presented in \cite{Cao:2013gug}, where the general BAE for periodic (or anti-periodic) XYZ spin chain (with spin specified to $\frac{1}{2}$) is obtained. There are in general three $Q$-functions appearing in the TQ-relation respectively denoted as, 
\begin{equation}
Q_1(u)=\prod_{j=1}^M\frac{\sigma(u-\mu_j)}{\sigma(\eta)},\quad Q_2(u)=\prod_{j=1}^M\frac{\sigma(u-\nu_j)}{\sigma(\eta)},\quad Q(u)=\prod_{j=1}^{M_1}\frac{\sigma(u-\lambda_j)}{\sigma(\eta)}.    
\end{equation}
The number of zeroes of these $Q$-functions are respectively parameterized by $M$, $M$ and $M_1$, and we further introduce a new integer number $m$ satisfying 
\begin{equation}
    L+m=2M+M_1.
\end{equation}
The most general form of the T-Q relation is given by 
\begin{align}
    \Lambda(u)=e^{2\pi i l_1 u+i\phi}a(u)\frac{Q_1(u-\eta)Q(u-\eta)}{Q_2(u)Q(u)}+e^{-2\pi i l_1 (u+\eta)-i\phi}d(u)\frac{Q_2(u+\eta)Q(u+\eta)}{Q_1(u)Q(u)}\cr
+c\frac{\sigma^m(u+\frac{\eta}{2})a(u)d(u)}{\sigma^m(\eta)Q_1(u)Q_2(u)Q(u)},\label{TQ-XYZ}
\end{align}
where $a(u)$ and $d(u)$ are the elliptic version of the eigenvalue of ${\cal A}$ and ${\cal D}$ acting on the ground state, and they are given by 
\ba
a(u)=\prod_{i=1}^L\frac{\sigma(u-\theta_i+\eta)}{\sigma(\eta)},\quad d(u)=\prod_{i=1}^L\frac{\sigma(u-\theta_i)}{\sigma(\eta)}.
\ea
In addition to the condition that $\Lambda(u)$ needs to be an entire function, which gives the BAEs, there are also two conditions from the quasi-periodicity of $\Lambda$ (similar to the elliptic $\theta$-functions), 
\ba
\lt(\frac{L}{2}-M-M_1\rt)\eta-\sum_{i=1}^M(\mu_i-\nu_i)=l_1\tau+m_1,\label{cons-1}
\ea
for $^\exists l_1,m_1\in\mathbb{Z}$, and 
\ba
\frac{L}{2}\eta-\sum_{i=1}^L \theta_i+\sum_{i=1}^M(\mu_i+\nu_i)+\sum_{i=1}^{M_1}\lambda_i=m_2,
\ea
for $^\exists m_2\in\mathbb{Z}$. The BAEs read
\begin{align}
&ce^{2\pi i l_1(\mu_j+\eta)+i\phi}a(\mu_j)=-Q_2(\mu_j)Q_2(\mu_j+\eta)Q(\mu_j+\eta),\\
&ce^{-2\pi i l_1\nu_j-i\phi}d(\nu_j)=-Q_1(\nu_j)Q_1(\nu_j-\eta)Q(\nu_j-\eta),\\
&\frac{e^{2\pi il_1(2\lambda_j+\eta)+2\phi i}a(\lambda_j)}{d(\lambda_j)}+\frac{Q_2(\lambda_j)Q_2(\lambda_j+\eta)Q(\lambda_j+\eta)}{Q_1(\lambda_j)Q_1(\lambda_j-\eta)Q(\lambda_j-\eta)}\nn\\
&=-\frac{ce^{2\pi il_1(\lambda_j+\eta)+\phi i}a(\lambda_j)}{Q_1(\lambda_j)Q_1(\lambda_j-\eta)Q(\lambda_j-\eta)}.
\end{align}

We are interested in the special case that $c=0$. From the first two equations above, $\{\mu_i\}$ and $\{\nu_i\}$ are forced to be paired in the form either $\mu_j=\nu_k$ or $\mu_j=\nu_k-\eta$. Let us relabel the first $\bar{m}$ roots of $Q_1$ and $Q_2$ s.t. they satisfy $\mu_i=\nu_i$ for $i=1,\dots,\bar{m}$, and $\mu_j=\nu_j-\eta$ for $j=\bar{m}+1,\dots,M$, then  
\begin{equation}
    Q_1(u)=\prod_{i=1}^{\bar{m}}\frac{\sigma(u-\mu_i)}{\sigma(\eta)}\prod_{j=\bar{m}+1}^{M}\frac{\sigma(u-\mu_j)}{\sigma(\eta)},\quad Q_2(u)=\prod_{i=1}^{\bar{m}}\frac{\sigma(u-\mu_i)}{\sigma(\eta)}\prod_{j=\bar{m}+1}^{M}\frac{\sigma(u-\mu_j-\eta)}{\sigma(\eta)}.
\end{equation}
One can then easily see that the TQ-relation is simplified to 
\begin{equation}
    \Lambda(u)=e^{2\pi i l_1 u+i\phi}a(u)\frac{\bar{Q}(u-\eta)}{\bar{Q}(u)}+e^{-2\pi i l_1 (u+\eta)-i\phi}d(u)\frac{\bar{Q}(u+\eta)}{\bar{Q}(u)},
\end{equation}
where 
\begin{equation}
    \bar{Q}(u):=\prod_{i=1}^{\bar{m}}\frac{\sigma(u-\mu_i)}{\sigma(\eta)}\prod_{j=1}^{M_1}\frac{\sigma(u-\lambda_j)}{\sigma(\eta)}=:\prod_{j=1}^{\bar{M}}\frac{\sigma(u-u_j)}{\sigma(\eta)},
\end{equation}
with the total number of roots $u_j$'s in $\bar{Q}$ denoted as $\bar{M}$, i.e. $\bar{M}=M_1+\bar{m}$. 
The constraint \eqref{cons-1} becomes 
\begin{equation}
    \lt(\frac{L}{2}-\bar{m}-M_1\rt)\eta=l_1\tau+m_1.\label{cons-p1}
\end{equation}
The BAE is also simplified to 
\begin{equation}
    \prod_{i=1}^L\frac{\sigma(u_j-\theta_i+\eta)}{\sigma(u_j-\theta_i)}=-e^{-4\pi i l_1 u_j-2i\phi}\frac{\bar{Q}(u_j+\eta)}{\bar{Q}(u_j-\eta)}.\label{BAE-XYZ}
\end{equation}

When generic $\eta$ and $\tau$ are considered, the only way to meet the \eqref{cons-p1} is to require 
\begin{equation}
    l_1=m_1=0,\quad L=2(M_1+\bar{m}).
\end{equation}
Therefore the BAE in this case becomes 
\begin{equation}
    \prod_{i=1}^L\frac{\sigma(u_j-\theta_i+\eta)}{\sigma(u_j-\theta_i)}=-e^{-2i\phi}\frac{\bar{Q}(u_j+\eta)}{\bar{Q}(u_j-\eta)},
\end{equation}
and it is very similar to the elliptic uplift of the XXZ BAE. 
In this case the number of sites, however, must be even, and is twice the number of roots in $\bar{Q}$, $\bar{M}=M_1+\bar{m}$. There is also a non-trivial restriction on $\phi$ called the selection rule, coming from the identity \cite{Cao:2013gug},  
\begin{equation}
    \prod_{i=1}^L\Lambda(\theta_i)=\prod_{i=1}^L a(\theta_i).
\end{equation}
In the homogeneous limit $\theta_i\rightarrow 0$, $a(0)=1$ and thus the selection rule becomes 
\begin{equation}
    e^{i\phi}\prod_{j=1}^{\bar{M}}\frac{\sigma(u_j+\eta)}{\sigma(u_j)}=e^{\frac{2\pi ik}{L}},\quad k=1,2,\dots,L.
\end{equation}
One can see from the above restriction and also from the numerical results presented in \cite{Cao:2013gug} that the emergent parameter $\phi$ is quantized (and related to $\eta$) due to the choice of $k$. The BAE of this XYZ model is simply an elliptic uplift of that of the XXZ model \eqref{BAE-p} (up to a shift of $u_j\rightarrow u_j-\eta/2$), and can be identified with $U(k)$ gauge theory with even $N_f$ and $2k=N_f$, whose vacua equation is given by \eqref{4d-vacua}. When $\phi$ is a constant here, however, the phase factor $E$ defined in \eqref{def-Ephi} must also be a constant, and we need to freeze the U(1) part of $\Phi_{U(1)}$ (or simply consider an SU($k$) gauge theory) and need to require the coefficient of $\Phi_i$ in $E$ to be zero. Together with the modularity condition \eqref{modularity}, one can see that $m_{adj}$, which corresponds to $\eta$ in the spin chain, takes some specific rational value and we shall turn to the following case instead. 

Let $\eta$ to take some specific values characterized by non-zero integers $l_1$ and $m_1$ as in \eqref{cons-p1}, and we introduce a new integer number $\bar{L}:=L-2\bar{M}$ to have 
\begin{equation}
    \bar{L}\eta=2l_1\tau+2m_1.
\end{equation}
The value of $\phi$ is found to be \cite{Samaj:2013yva}
\begin{equation}
    \phi(u)=\frac{2\pi}{\bar{L}}\lt[m+4l_1\sum_{k=1}^{\bar{M}}\lt(u_k+\frac{\eta}{2}\rt)+2l_1\bar{L}\lt(u+\frac{\eta}{2}\rt)\rt].
\end{equation}
By comparing the vacua equation \eqref{4d-vacua} with the BAE \eqref{BAE-XYZ}, in particular the coefficients of $u_i\leftrightarrow \Phi_i$ and $\sum_j u_j\leftrightarrow \Phi_{U(1)}$ in the phase $\phi$ and $E$ in \eqref{def-Ephi}, one can solve all the constraints, 
\begin{align}
    &-\frac{4\pi i}{\varsigma}\leftrightarrow -2i\frac{2\pi}{\bar{L}}4\ell_1,\\
    &-\frac{2\pi i}{\varsigma}N_f+\frac{2\pi i}{\varsigma}\lt(\sum_fm_f+\sum_{\bar{f}}\bar{m}_{\bar{f}}\rt)+\frac{4\pi i}{\varsigma}km_{adj}\leftrightarrow -12\pi i\ell_1,
\end{align}
to have
\begin{equation}
    \bar{M}=\frac{L}{2(3-\eta)},
\end{equation}
i.e. the number of magnons is still restricted to some special value. In the above computation, we combined the modularity condition \eqref{modularity} together with the dictionary, 
\begin{align}
    k\leftrightarrow \bar{M},\quad m_{adj}\leftrightarrow \eta, \quad N_f\leftrightarrow L,\quad \varsigma\leftrightarrow \tau. 
\end{align}
We see here that a different feature of the Bethe/Gauge correspondence in 4d from its lower-dimensional versions is that the rank of the gauge group is restricted according to the number of flavors in the gauge theory.

\section{Bethe/Gauge correspondence to open spin chains}\label{s:bethe-open}

In this section, we turn to the Bethe/Gauge correspondence to open spin chains whose details were provided in \cite{Kimura:2020bed}, and we try to generalize the results there to higher-rank cases together with off-diagonal boundary conditions. 

To construct an open integrable quantum system, one needs to find one solution to the boundary YBE,
\begin{align}
    R(u-v)K^-_1(u)R(u+v)K^-_2(v)=K^-_2(v)R(u+v)K^-_1(u)R(u-v),\label{bYBE}
\end{align}
to specify the boundary condition at each end. We take such a solution $K^-(u)$ and a dual solution constructed as 
\begin{equation}
    K^+(u)=\bar{K}^-\lt(-u-\frac{h\eta}{2}\rt)M,
\end{equation}
from another solution $\bar{K}^-$ to \eqref{bYBE}, where $h$ is a parameter and $M$ is a matrix acting on one site associated to the crossing-unitary relation, 
\begin{equation}
    R^{t_1}_{12}(u)M_1R^{t_1}_{21}(-u-h\eta)M_1^{-1}\propto {\rm id}.
\end{equation}
The transfer matrix of the open spin chain is constructed as 
\begin{equation}
t(u)={\rm tr}_0K_0^-(u){\bf T}_{0}(u)K_0^+(u){\bf T}^{-1}_{0}(-u),
\end{equation}
with the monodromy matrix defined by 
\begin{equation}
{\bf T}_0(u):=L_{0L}(u-\theta_L)L_{0(L-1)}(u-\theta_{L-1})\dots L_{02}(u-\theta_2)L_{01}(u-\theta_1).
\end{equation}
When both $K^\pm$ are diagonal, the techniques to derive the BAE were developed in \cite{Sklyanin:1988yz}, and for more general boundary operators, the off-diagonal BAE method proposed in \cite{Cao:2013nza} can be applied to diagonalize the transfer matrix. In this section, we compare the BAEs obtained so with the vacua equations of various kinds of gauge theories in 3d and 4d to propose potential versions of the Bethe/Gauge correspondence.

\subsection{3d and 2d version corresponding to rank-1 open chains}\label{s:BG-KZ}

The correspondence proposed in \cite{Kimura:2020bed} focused on the case of open spin chains with diagonal boundary operators $K^\pm$. Each of the boundary is parameterized by one parameter $\xi^\pm$, 
\begin{align}
    &K^-(u)=\lt(\begin{array}{cc}
      \sinh(u+\xi^-)e^{u}   &  \\
         & -\sinh(u-\xi^-)e^{-u}
    \end{array}\rt),\cr
    &K^+(u)=\lt(\begin{array}{cc}
      \sinh(u+\xi^+-\frac{\eta}{2})e^{u}   &  \\
         & -\sinh(u-\xi^++\frac{\eta}{2})e^{-u}
    \end{array}\rt).
\end{align}
The BAE in this case reads,
\begin{align}
    \frac{\sinh\lt(u_i+\xi_+-\frac{\eta}{2}\rt)\sinh\lt(u_i-\frac{\eta}{2}+\xi_-\rt)\delta_+(u_i)\delta_-(-u_i)}{\sinh\lt(u_i-\xi_++\frac{\eta}{2}\rt)\sinh\lt(u_i+\frac{\eta}{2}-\xi_-\rt)\delta_+(-u_i)\delta_-(u_i)}\cr
    \times\prod_{j\neq i}^m \frac{\sinh(u_j-u_i+\eta)\sinh(u_i+u_j-\eta)}{\sinh(u_j-u_i-\eta)\sinh(u_j+u_i+\eta)}=1,
\end{align}
where 
\begin{align}
    \delta_+(u)=\prod_{a=1}^L\sinh(u+\eta/2+\eta s_a-\theta_a),\cr
    \delta_-(u)=\prod_{a=1}^L\sinh(u+\eta/2-\eta s_a-\theta_a).
\end{align}
The BAEs in such open spin chains are proposed in \cite{Kimura:2020bed} to match with the vacua equations in SO and Sp-type gauge theories. 

In SO($2k+\delta$) theory with $N_f$ flavors in the vector representation, the vacua equation derived from \eqref{vacua-eq} is given by 
\begin{equation}
    \frac{[\sigma_i-\beta_2\tilde{c}]^\delta}{[\sigma_i+\beta_2\tilde{c}]^\delta}\prod_{j\neq i}\frac{[\sigma_i\pm \sigma_j-\beta_2\tilde{c}]}{[-\sigma_i\pm \sigma_j-\beta_2\tilde{c}]}\prod_{a=1}^{N_f}\frac{[\sigma_i-m_a-\beta_2\tilde{c}]}{[-\sigma_i-m_a-\beta_2\tilde{c}]}=1.
\end{equation}
We note here that one may further add a contribution from the boundary Chern-Simons term, 
\begin{equation}
    S_{bCS}=\frac{\kappa}{4\beta}{\rm tr}\sigma^2=\frac{\kappa}{4\beta}\sum_i\sigma_i^2, \label{s-bcs}
\end{equation}
to introduce an exponential term $e^{\frac{i\kappa\beta_2}{2\beta}\sigma_i}$. It is straightforward to read off the map mimicking \eqref{U-map}, 
\begin{align}
    &m\leftrightarrow k,\quad 2L\leftrightarrow N_f,\quad u_j\leftrightarrow i\sigma_j/2,\quad \eta\leftrightarrow i\beta_2\tilde{c}/2,\cr
    &\lt\{\eta s_a+\frac{\eta}{2}-\theta_a,\eta s_a-\frac{\eta}{2}+\theta_a\rt\}_{a=1}^L\leftrightarrow \{-m_f-\beta_2\tilde{c}\}_{f=1}^{N_f},\label{SO-map}
\end{align}
and the boundary parameters should be chosen properly to reproduce the root structure of the gauge group under consideration. In the case of $\delta=0$, one can choose 
\begin{equation}
    \xi_\pm=\infty,\label{SO2N-bc}
\end{equation}
to realize the vacua equation, and for $\delta=1$, it can be realized by e.g. 
\begin{equation}
    \xi_+=\infty, \quad \xi_-=-\frac{\eta}{2}.\label{SO2N1-bc}
\end{equation}
An interesting feature\footnote{We thank Prof. Yunfeng Jiang for pointing out this property of the BAE that is not often stressed in the literature. } here is that the same BAE can be realized by different choices of the boundary parameters, for example for $\delta=0$, one can alternatively impose $\xi_+=-\xi_-+\eta$ with a free parameter $\xi_-$ unspecified. The special choice $\xi_+=\frac{\eta}{2}$ also does not give any contributing factor to the BAE, so corresponding to the case $\delta=1$, it gives rise to the same BAE with 
\begin{equation}
    \xi_+=\frac{\eta}{2}, \quad \xi_-=-\frac{\eta}{2}.\label{SO2N1-bca}
\end{equation}

The vacua equation of the Sp gauge theory is given by 
\begin{equation}
    \frac{[2\sigma_i-\beta_2\tilde{c}]^2}{[2\sigma_i+\beta_2\tilde{c}]^2}\prod_{j\neq i}\frac{[\sigma_i\pm \sigma_j-\beta_2\tilde{c}]}{[-\sigma_i\pm \sigma_j-\beta_2\tilde{c}]}\prod_{a=1}^{N_f}\frac{[\sigma_i-m_a-\beta_2\tilde{c}]}{[-\sigma_i-m_a-\beta_2\tilde{c}]}=1.\label{vacua-Sp}
\end{equation}
One can use 
\begin{equation}
    \sin(2x)=2\sin(x)\cos(x)=2\sin(x)\sin\lt(x+\frac{\pi}{2}\rt),
\end{equation}
to convert 
\begin{equation}
    [2\sigma_i-\beta_2\tilde{c}]=[\sigma_i-\beta_2\tilde{c}/2][\sigma_i-\beta_2\tilde{c}/2+\pi].
\end{equation}
Four boundary parameters are needed in this frame of realizing the Bethe/Gauge correspondence, so it cannot be achieved for 3d gauge theories but can alternatively be realized in the 2d limit $\beta_2\rightarrow 0$, where $[\sigma_i-\beta_2\tilde{c}/2]\rightarrow (\sigma_i-\beta_2\tilde{c}/2)$ and $[\sigma_i-\beta_2\tilde{c}/2+\pi]\rightarrow 1$ \cite{Kimura:2020bed}. One can then easily see that the choice
\begin{equation}
    \xi_+=0,\quad \xi_-=0,
\end{equation}
realizes the vacua equation of 2d Sp($k$) theory.

\subsubsection{How about spin chains with off-diagonal boundaries?}\label{s:off-diag}

Since the naive Bethe/Gauge correspondence does not work for Sp($k$) gauge theories in 3d by using the diagonal open boundaries, one may then ask what if we use the most general off-diagonal open boundaries, where more boundary parameters are available? 

One can consider a general solution to \eqref{bYBE} with off-diagonal entries parameterized by three boundary parameters $(\alpha_-,\beta_-,\theta_-)$, 
\begin{align}
    K^-_{11}(u)=2(\sinh(\alpha_-)\cosh(\beta_-)\cosh(u)+\cosh(\alpha_-)\sinh(\beta_-)\sinh(u)),\\
    K^-_{22}(u)=2(\sinh(\alpha_-)\cosh(\beta_-)\cosh(u)-\cosh(\alpha_-)\sinh(\beta_-)\sinh(u)),\\
    K^-_{12}(u)=e^{\theta_-}\sinh(2u),\quad K^-_{21}(u)=e^{-\theta_-}\sinh(2u),
\end{align}
and also a dual solution characterized by another set of boundary parameters $(\alpha_+,\beta_+,\theta_+)$. The BAE (in the case of spin $1/2$) is derived in \cite{Zhang:2014ria} again with the off-diagonal BAE method, 
\begin{align}
    \frac{\sinh(2u_i+\eta)\sinh(u_i-\alpha_\pm-\eta/2)\cosh(u_i-\beta_\pm-\eta/2)}{\sinh(-2u_i+\eta)\sinh(-u_i-\alpha_\pm-\eta/2)\cosh(-u_i-\beta_\pm-\eta/2)}\cr
    \times\prod_{l=1}^L\frac{\sinh(u_i-\theta_l+\eta/2)\sinh(u_i+\theta_l+\eta/2)}{\sinh(-u_i-\theta_l+\eta/2)\sinh(-u_i+\theta_l+\eta/2)}\cr
    \times\prod_{j=1}^L\frac{\sinh(u_i-u_j-\eta)\sinh(u_i+u_j-\eta)}{\sinh(u_i-u_j+\eta)\sinh(u_i+u_j+\eta)}=1,
\end{align}
where a constraint is necessary on six boundary parameters, 
\begin{equation}
    \cosh(\alpha_++\alpha_-+\beta_++\beta_-+(L+1)\eta)-\cosh(\theta_--\theta_+)=0,
\end{equation}
to kill the third term (to set $c=0$) in the general TQ-relation as in \eqref{TQ-XYZ}. To realize the vacua equation \eqref{vacua-Sp}, one needs to reproduce a factor 
\begin{equation}
    \frac{\sinh^2(2u_i-\eta)}{\sinh^2(2u_i+\eta)},
\end{equation}
with the boundary parameters in the BAE. Unfortunately, an inverse factor already exists in the BAE and four boundary parameters $\alpha_\pm$ and $\beta_\pm$ cannot fully recover the above factor. We have to conclude that 3d Sp theories do not have Bethe dual spin chains in our approach. We note that in \cite{Ding:2023auy}, a rather new way to reproduce the vacua equation from the spin chain BAEs was applied, but we will not pursue further in this direction.

\subsection{Bethe/Gauge correspondence for $A_N$ spin chains}

As originally mentioned briefly in \cite{Nekrasov:2009ui}, the Bethe/Gauge correspondence can be generalized to that between the quiver gauge theories and spin chains of ADE-type. We only consider the case of the linear quivers in this article. 

The simplest example is given by the closed $A_N$ spin chain whose BAE follows from the nested Bethe ansatz method and is given by 
\begin{align}
    &e^{i\theta_k}\prod_{j\neq i}^{r_k}\frac{\sinh(u^{(k)}_i-u^{(k)}_j-\eta)}{\sinh(u^{(k)}_i-u^{(k)}_j+\eta)}\cr
    &=\prod_{j=1}^{r_{k+1}}\frac{\sinh(u^{(k)}_i-u^{(k+1)}_j-\eta/2)}{\sinh(u^{(k)}_i-u^{(k+1)}_j+\eta/2)}\prod_{j=1}^{r_{k-1}}\frac{\sinh(u^{(k)}_i-u^{(k-1)}_j-\eta/2)}{\sinh(u^{(k)}_i-u^{(k-1)}_j+\eta/2)}.
\end{align}
A similar vacua equation can be derived from a quiver gauge theory with the following quiver structure. 
\begin{align}
    \begin{tikzpicture}
    \draw[ultra thick,->] (0,0) to [out=45,in=180] (1,0.5);
    \draw[ultra thick] (1,0.5) to [out=0,in=135] (2,0);
    \draw[ultra thick,->] (2,0) to [out=-135,in=0] (1,-0.5);
    \draw[ultra thick] (1,-0.5) to [out=180,in=-45] (0,0);
    \draw[ultra thick,->] (-2,0) to [out=45,in=180] (-1,0.5);
    \draw[ultra thick] (-1,0.5) to [out=0,in=135] (0,0);
    \draw[ultra thick,->] (0,0) to [out=-135,in=0] (-1,-0.5);
    \draw[ultra thick] (-1,-0.5) to [out=180,in=-45] (-2,0);
    \draw[thick] (3,0)--(2,0);
    \draw[thick] (-3,0)--(-2,0);
        \draw[fill=yellow,thick] (0,0) circle (0.7); 
        \draw[fill=yellow,thick] (2,0) circle (0.7);
        \draw[fill=yellow,thick] (-2,0) circle (0.7);
        \node at (0,0) {U($r_k$)};
        \node at (2,0) {U($r_{k+1}$)};
        \node at (-2,0) {U($r_{k-1}$)};
        \node at (3.5,0) {\dots};
        \node at (-3.5,0) {\dots};
    \end{tikzpicture}
\end{align}
In the nested Bethe ansatz approach, the full $\mathfrak{su}(N+1)$ algebra is divided into layers of $\mathfrak{su}(2)$ sub-algebras, and one of the $\mathfrak{su}(2)$ layer, usually chosen to be the last layer, carries the information of the representation $s_a$ (which usually corresponds to the $s_a$-th symmetric representation of $\mathfrak{su}(N+1)$ at the $a$-th site) and the inhomogeneous parameter $\theta_a$ at each site of the lattice chain. Such a layer is also often called the momentum carrying node in the literature, as each layer of $\mathfrak{su}(2)$ corresponds to a node in the quiver diagram. For $\mathfrak{su}(N+1)$ spin chain, the quiver takes exactly the same form as the Dynkin diagram, but corresponding to the momentum carrying node, matter chiral multiplets are attached in the quiver diagram: 
\begin{align}
    \begin{tikzpicture}
    \draw[thick] (-3,0)--(-2,0);
    \draw[ultra thick,->] (-2,0)--(-1,1.5/2);
    \draw[ultra thick] (0,1.5)--(-1,1.5/2);
    \draw[ultra thick] (-2,0)--(-1,-1.5/2);
    \draw[ultra thick,->] (0,-1.5)--(-1,-1.5/2);
        \draw[fill=yellow,thick] (-2,0) circle (0.7);
        \draw[fill=yellow,thick] (-0.5,1) rectangle (0.5,2);
        \draw[fill=yellow,thick] (-0.5,-1) rectangle (0.5,-2);
        \node at (-2,0) {U($r_{N}$)};
        \node at (-3.5,0) {\dots};
        \node at (0,1.5) {$L$};
        \node at (0,-1.5) {$L$};
    \end{tikzpicture}
\end{align}

We are more interested in the Bethe/Gauge correspondence with open spin chains, which are expected to be related to SO or Sp gauge theories. In such gauge theories, the representations are real and we do not need to pair the fundamental matters with anti-fundamental ones. We simply consider the following linear quiver as dual candidates for the open spin chain, 
\begin{align}
    \begin{tikzpicture}
    \draw[ultra thick] (0,0)--(2,0);
    \draw[ultra thick] (0,0)--(-2,0);
    \draw[ultra thick] (2,0)--(4,0);
    \draw[thick] (-2,0)--(-3,0);
        \draw[fill=yellow,thick] (0,0) circle (0.7);
        \draw[fill=yellow,thick] (2,0) circle (0.7);
        \draw[fill=yellow,thick] (-2,0) circle (0.7);
        \draw[fill=yellow,thick] (3.5,0.5) rectangle (4.5,-0.5);
        \node at (-2,0) {$G_{N-2}$};
        \node at (0,0) {$G_{N-1}$};
        \node at (2,0) {$G_{N}$};
        \node at (4,0) {$L$};
        \node at (-3,0) [left] {\dots};
    \end{tikzpicture}
    \label{linear-quiver}
\end{align}
where all the gauge group $G_i$'s are taken to the BCD-type. 

\subsubsection{Bethe/Gauge correspondence to open $A_N$ XXZ spin chain with diagonal boundaries}

Let us briefly describe the system of $A_N$ spin chain first and summarize the results available in the literature on its BAEs, then we will compare the BAEs with the corresponding vacua equations. 

The generalization to $A_N$ spin chain can rather be easily understood in the XXX chain, as the R-matrix is of the form 
\begin{equation}
    R_{ab}(u)=u\mathbb{I}_{ab}+i\mathbb{P}_{ab},
\end{equation}
and it can be uplift to an arbitrary $N\times N$ matrix by using the expression of the permutation operator exchanging $N$-dim vectors, 
\begin{equation}
    \mathbb{P}=\sum_{i,j=1}^NE_{ij}\otimes E_{ji},\quad (E_{ij})_{kl}=\delta_{i,k}\delta_{j,l}.
\end{equation}
Then we simply need to solve for the diagonal solutions $K(u)={\rm diag}(k_1,k_2,\dots,k_n)$ to the BYBE \eqref{bYBE}. It is then not hard to find each component as $k_i=(u+\xi)$ or $k_i=(-u+\xi)$.

A similar formula for the general form of the diagonal K-matrices $K^-={\rm diag}(K^-_a)$ for $A_N$-type higher-rank open XXZ spin chain has been worked out in \cite{deVega:1992zd} as 
\begin{align}
    K^-_a(u)=k\sinh(u+\xi)e^u,\quad 1\leq a\leq l_-,\cr
    K^-_a(u)=k\sinh(\xi-u)e^{-u},\quad l_-+1\leq a\leq N.
\end{align}
By defining a diagonal matrix $M$ with components $M_{ab}=\delta_{ab}e^{(N-2a+1)\gamma}$ with $\gamma:=\frac{2\eta}{N}$, the dual solution $K^+$ can be found via 
\begin{equation}
    K^+(u)=K^-(-u-\eta)^tM,
\end{equation}
which in the current case is also a diagonal matrix. 

The BAE for the general diagonal higher-rank $A_{N-1}$ open spin chains has been worked out in \cite{deVega:1994sb}. We denote the Bethe roots associated to the $k$-th $\mathfrak{su}(2)$ layer in the nesting process as $\{u_i^{(k)}\}_{i=1}^{r_k}$, then the BAE reads 
\begin{align}
    h^{(k)}(u_i^{(k)})\prod_{j\neq i}^{r_k}\frac{\sinh(u_i^{(k)}+u_j^{(k)}+(k-1)\gamma)\sinh(u_i^{(k)}-u_j^{(k)}-\gamma)}{\sinh(u_i^{(k)}+u_j^{(k)}+(k+1)\gamma)\sinh(u_i^{(k)}-u_j^{(k)}+\gamma)}\cr
    =\prod_{j=1}^{r_{k+1}}\frac{\sinh(u_i^{(k)}+u_j^{(k+1)}+k\gamma)\sinh(u_i^{(k)}-u_j^{(k+1)}-\gamma)}{\sinh(u_i^{(k)}+u_j^{(k+1)}+(k+1)\gamma)\sinh(u_i^{(k)}-u_j^{(k+1)})}\cr
    \times\prod_{j=1}^{r_{k-1}}\frac{\sinh(u_i^{(k)}+u_j^{(k-1)}+(k-1)\gamma)\sinh(u_i^{(k)}-u_j^{(k-1)})}{\sinh(u_i^{(k)}+u_j^{(k-1)}+k\gamma)\sinh(u_i^{(k)}-u_j^{(k-1)}+\gamma)},\label{general-BAE-open}
\end{align}
where if $l_+\neq l_-$, 
\begin{align}
    h^{(l_-)}(u_i^{(l_-)})=\frac{\sinh(\xi_--u_i^{(l_-)})}{\sinh(\xi_-+u_i^{(l_-)}+l_-\gamma)}e^{2u_i^{(l_-)}+l_-\gamma},\\
    h^{(l_+)}(u_i^{(l_+)})=\frac{\sinh(\xi_++u_i^{(l_+)}-(N-l_+)\gamma)}{\sinh(\xi_+-u_i^{(l_+)}-N\gamma)}e^{-2u_i^{(l_+)}-l_+\gamma},
\end{align}
and if $l_+=l_-=l$, 
\begin{equation}
    h^{(l)}(u_i^{(l)})=h^{(l_-)}(u_i^{(l_-)})h^{(l_+)}(u_i^{(l_+)}),
\end{equation}
while for other $l$, $h^{(l)}(u)=1$. 

To compare with the vacua equation of the quiver gauge theory, we perform a shift to replace,
\begin{equation}
    u^{(k)}_i\rightarrow u^{(k)}_i-\frac{k}{2}\gamma,
\end{equation}
then the general BAE \eqref{general-BAE-open} becomes 
\begin{align}
    \tilde{h}^{(k)}(u_i^{(k)})\prod_{j\neq i}^{r_k}\frac{\sinh(u_i^{(k)}+u_j^{(k)}-\gamma)\sinh(u_i^{(k)}-u_j^{(k)}-\gamma)}{\sinh(u_i^{(k)}+u_j^{(k)}+\gamma)\sinh(u_i^{(k)}-u_j^{(k)}+\gamma)}\cr
    =\prod_{j=1}^{r_{k+1}}\frac{\sinh(u_i^{(k)}+u_j^{(k+1)}-\gamma/2)\sinh(u_i^{(k)}-u_j^{(k+1)}-\gamma/2)}{\sinh(u_i^{(k)}+u_j^{(k+1)}+\gamma/2)\sinh(u_i^{(k)}-u_j^{(k+1)}+\gamma/2)}\cr
    \times\prod_{j=1}^{r_{k-1}}\frac{\sinh(u_i^{(k)}+u_j^{(k-1)}-\gamma/2)\sinh(u_i^{(k)}-u_j^{(k-1)}-\gamma/2)}{\sinh(u_i^{(k)}+u_j^{(k-1)}+\gamma/2)\sinh(u_i^{(k)}-u_j^{(k-1)}+\gamma/2)},\label{BAE-open-An}
\end{align}
with 
\begin{align}
    &\tilde{h}^{(l_-)}(u_i^{(l_-)})=\frac{\sinh(\tilde{\xi}_--u_i^{(l_-)})}{\sinh(\tilde{\xi}_-+u_i^{(l_-)})}e^{2u_i^{(l_-)}},\\
    &\tilde{h}^{(l_+)}(u_i^{(l_+)})=\frac{\sinh(\tilde{\xi}_++u_i^{(l_+)})}{\sinh(\tilde{\xi}_+-u_i^{(l_+)})}e^{-2u_i^{(l_+)}}.
\end{align}
where we introduced the notation $\tilde{\xi}_-=\xi_-+l_-\gamma/2$ and $\tilde{\xi}_+=\xi_+-(N-l_+/2)\gamma$ to further simplify the expressions.

We note here that the open spin chain used in the Bethe/Gauge correspondence described in section \ref{s:BG-KZ} corresponds to the case $l_+=l_-=1$. In the case of rank-1 spin chain, we could have also considered the case $l_+\neq l_-$, then if one of $l_\pm$ is equal to one, an effective contribution from the Chern-Simons term \eqref{s-bcs} is introduced, with the CS level restricted to 
\begin{equation}
    \pm 2\leftrightarrow \frac{\beta_2}{2\beta}\kappa.\label{open-CS}
\end{equation}
We also note that in the XXX limit, the exponential terms go to $1$, and taking $l_+\neq l_-$ with one of them equal to one is equivalent to putting $\xi_+=\infty$ or $\xi_-=\infty$. We see that SO theories can alternatively be realized by such a choice. 

The vacua equation of the linear quiver gauge theory (of the shape \eqref{linear-quiver}) with BCD-type gauge groups at the $\alpha$-th node is given by 
\begin{align}
    &\frac{\sin^{2\delta_{\alpha,Sp}}(2\sigma^{(\alpha)}_i-\beta_2\tilde{c}_\alpha)}{\sin^{2\delta_{\alpha,Sp}}(2\sigma^{(\alpha)}_i+\beta_2\tilde{c}_\alpha)}\prod_{j\neq i}^{N_\alpha}\frac{\sin(\sigma^{(\alpha)}_i\pm \sigma^{(\alpha)}_j-\beta_2\tilde{c}_\alpha)}{\sin(-\sigma^{(\alpha)}_i\pm \sigma^{(\alpha)}_j-\beta_2\tilde{c}_\alpha)}\prod_{j=1}^{N_{\alpha-1}}\frac{\sin(\sigma^{(\alpha)}_i\pm\sigma^{(\alpha-1)}_j-m_{bfd})}{\sin(-\sigma^{(\alpha)}_i\pm\sigma^{(\alpha-1)}_j-m_{bfd})}\cr
&\times\prod_{k=1}^{N_{\alpha+1}}\frac{\sin(\sigma^{(\alpha)}_i\pm\sigma^{(\alpha+1)}_k-m_{bfd})}{\sin(-\sigma^{(\alpha)}_i\pm\sigma^{(\alpha+1)}_k-m_{bfd})}\times\frac{\sin^{\delta_{\alpha-1}}(\sigma^{(\alpha)}_i-m_{bfd})}{\sin^{\delta_{\alpha-1}}(-\sigma^{(\alpha)}_i-m_{bfd})}\frac{\sin^{\delta_{\alpha+1}}(\sigma^{(\alpha)}_i-m_{bfd})}{\sin^{\delta_{\alpha+1}}(-\sigma^{(\alpha)}_i-m_{bfd})}\cr
&\times\frac{\sin^{\delta_\alpha}(\sigma^{(\alpha)}_i-\beta_2\tilde{c}_\alpha)}{\sin^{\delta_\alpha}(\sigma^{(\alpha)}_i+\beta_2\tilde{c}_\alpha)}=1,
\end{align}
where if the $\alpha$-th gauge group is Sp-type, then $\delta_{\alpha,Sp}=1$ and $\delta_\alpha=0$, otherwise $\delta_{Sp}=0$ and the gauge group will be taken to SO($2N+\delta_\alpha$). 

To match the BAE \eqref{BAE-open-An} with the vacua equation, we note that to realize an SO($2N$)-node in the quiver, no $\tilde{h}^{(i)}$ factor is needed, while each SO($2N+1$)-node requires at least one such factor. Sp-type nodes again can only be reproduced in the 2d limit with two $\tilde{h}^{(i)}$ factors at the same node, i.e. $l_+=l_-$. From the above observation, we conclude that the BAEs of the open $A_N$ XXZ spin chains with diagonal boundaries can only reproduce the vacua equations of quiver gauge theories of the following types. 
\begin{itemize}
    \item 3d quiver gauge theories with only SO($2k_\alpha$)-type gauge nodes.  
    \item 3d quiver gauge theories with one SO($2k_\beta+1$) gauge node and other nodes are all of SO($2k_\alpha$)-type. 
    \item 3d quiver gauge theories with two SO($2k_\beta+1$)-type gauge nodes (and opposite boundary Chern-Simons level \eqref{open-CS}), and the remaining nodes are of SO($2k_\alpha$)-type.
\end{itemize}
In the XXX limit, in addition to the corresponding quiver gauge theories listed above reduced to 2d, one can also realize 
\begin{itemize}
\item 2d quiver gauge theories with one Sp($k_\beta$) gauge node and the remaining nodes are of SO($2k_\alpha$)-type.
\end{itemize}

\subsubsection{Bethe/Gauge correspondence for open $A_N$ XXX spin chain with general boundaries}

When we focus on the XXX open spin chains, the general boundaries can be considered and the BAEs of such spin chains have been studied in \cite{Cao:2013cja}. The boundary $K$-matrix is constructed from the fusion techniques \cite{deVega:1993xi,Kulish:1995ns}, and the BAEs are also obtained in a recursive manner. To write down the BAEs of such a spin chain system, we first decompose $N+1$ into two sets of integers, $(p,q)$ and $(\bar{p},\bar{q})$ s.t. $p+q=N+1$ and $\bar{p}+\bar{q}=N+1$. Then we introduce $N+1$ $K^{(l)}$-functions for $l=1,2,\dots,N+1$, satisfying 
\begin{equation}
    K^{(l)}(u)=K^{(l+1)}(-u-l\eta)\ {\rm or}\ K^{(l)}(u)=K^{(l+1)}(u),\quad l=1,2,\dots,N,
\end{equation}
and the constraint,
\begin{align}
    \prod_{l=1}^{N+1}K^{(l)}(u-(l-1)\eta)=(-1)^{q+\bar{q}}\prod_{k=0}^{\bar{q}-1}\lt(-u+\frac{N-1}{2}\eta-\bar{\xi}-k\eta\rt)\cr
    \prod_{k=0}^{\bar{p}-1}\lt(-u+\frac{N-1}{2}\eta+\bar{\xi}-k\eta\rt)\prod_{k=0}^{q-1}(u-\xi-k\eta)\prod_{k=0}^{p-1}(u+\xi-k\eta).
\end{align}
The BAEs then read 
\begin{align}
    &\frac{u^{(1)}_j-\eta/2}{u^{(1)}_j+\eta/2}\frac{K^{(2)}(u^{(1)}_j-\eta/2)}{K^{(1)}(u^{(1)}_j-\eta/2)}\prod_{l=1}^L\frac{(u^{(1)}_j-\theta_l+\eta/2)(u^{(1)}_j+\theta_l+\eta/2)}{(u^{(1)}_j-\theta_l-\eta/2)(u^{(1)}_j+\theta_l-\eta/2)}\cr
    &\prod_{k=1}^{M_1}\frac{(u_j^{(1)}-u_k^{(1)}+\eta)(u_j^{(1)}+u_k^{(1)}+\eta)}{(u_j^{(1)}-u_k^{(1)}-\eta)(u_j^{(1)}+u_k^{(1)}-\eta)}\prod_{n=1}^{M_2}\frac{(u^{(1)}_j-u^{(2)}_n-\eta/2)(u^{(1)}_j+u^{(2)}_n-\eta/2)}{(u^{(1)}_j-u^{(2)}_n+\eta/2)(u^{(1)}_j+u^{(2)}_n+\eta/2)}=-1,\cr
\end{align}
and 
\begin{align}
    \frac{u^{(l)}_j+\eta/2}{u^{(l)}_j-\eta/2}\frac{K^{(l)}(u^{(l)}_j-l\eta/2)}{K^{(l+1)}(u^{(l)}_j-l\eta/2)}\prod_{k=1}^{M_{l-1}}\frac{(u_j^{(l)}-u_k^{(l-1)}+\eta/2)(u_j^{(l)}+u_k^{(l-1)}+\eta/2)}{(u_j^{(l)}-u_k^{(l-1)}-\eta/2)(u_j^{(l)}+u_k^{(l-1)}-\eta/2)}\cr
    \times \prod_{k=1}^{M_{l+1}}\frac{(u_j^{(l)}-u_k^{(l+1)}+\eta/2)(u_j^{(l)}+u_k^{(l+1)}+\eta/2)}{(u_j^{(l)}-u_k^{(l+1)}-\eta/2)(u_j^{(l)}+u_k^{(l+1)}-\eta/2)}\cr
    \times \prod_{k=1}^{M_l}\frac{(u_j^{(l)}-u_k^{(l)}-\eta)(u_j^{(l)}+u_k^{(l)}-\eta)}{(u_j^{(l)}-u_k^{(l)}+\eta)(u_j^{(l)}+u_k^{(l)}+\eta)}=-1.
\end{align}
Attached to each node, there is a boundary contribution $\frac{K^{(l)}(u^{(l)}_j-l\eta/2)}{K^{(l+1)}(u^{(l)}_j-l\eta/2)}$ in such models, and this indicates a potential realization of arbitrary number of SO($2k_l+1$) or even Sp($k_l$) nodes in the quiver gauge theories. 

Let us give more explicit expressions of the boundary contributions. 
For simplicity, we start from the case of the $A_2$ open spin chains to compare with the vacua equations in 2d gauge theories. When choosing $q=\bar{q}=1$, $p=\bar{p}=2$, we have e.g. 
\begin{align}
    &K^{(1)}(u)=(u+\xi)(-u+\eta/2+\bar{\xi}),\\
    &K^{(2)}(u)=(-u+\xi-\eta)(u+3\eta/2+\bar{\xi}),\\
    &K^{(3)}(u)=(-u+\xi-\eta)(u+3\eta/2+\bar{\xi}).
\end{align}
As the contribution from $\frac{K^{(2)}(u^{(2)}_j-2\eta/2)}{K^{(3)}(u^{(2)}_j-2\eta/2)}$ becomes trivial in this case, the second node is still restricted to an SO($2k$)-type. It was remarked in \cite{Cao:2013cja} that although the boundary contributions are not uniquely determined, it is believed that different choices may only give equivalent parameterizations of the system (as it is proved in the $A_1$ case in \cite{Bazhanov:1987zu,Nepomechie:2013ila}). In the case of $A_3$, one may choose $p=\bar{p}=3$ and $q=\bar{q}=1$ to have 
\begin{align}
    &K^{(1)}(u)=(u+\xi)(-u-\eta+\bar{\xi}),\\
    &K^{(2)}(u)=(-u+\xi-\eta)(u+\bar{\xi}),\\
    &K^{(3)}(u)=(u+\xi+\eta)(-u-2\eta+\bar{\xi}),\\
    &K^{(4)}(u)=(u+\xi+\eta)(-u-2\eta+\bar{\xi}).
\end{align}
This choice introduces two boundary contributions respectively at the first and the second node, 
\begin{align}
    \frac{K^{(1)}(u^{(1)}_j-\eta/2)}{K^{(2)}(u^{(1)}_j-\eta/2)}=\frac{(u^{(1)}_j+\xi-\eta/2)(-u^{(1)}_j+\bar{\xi}-\eta/2)}{(-u^{(1)}_j+\xi-\eta/2)(u^{(1)}_j+\bar{\xi}-\eta/2)},\\
    \frac{K^{(2)}(u^{(2)}_j-\eta)}{K^{(3)}(u^{(2)}_j-\eta)}=\frac{(-u^{(2)}_j+\xi)(u^{(2)}_j+\bar{\xi}-\eta)}{(u^{(2)}_j+\xi)(-u^{(2)}_j+\bar{\xi}+\eta)}.
\end{align}
Interestingly, by choosing 
\begin{equation}
    \xi=\frac{\eta}{2},\quad \bar{\xi}=\frac{3\eta}{2},
\end{equation}
one can realize the following quiver gauge theory, 
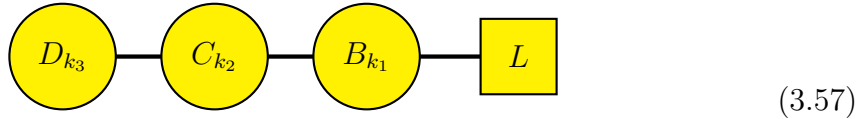
\begin{align}
    \begin{tikzpicture}
    \draw[ultra thick] (0,0)--(2,0);
    \draw[ultra thick] (0,0)--(-2,0);
    \draw[ultra thick] (2,0)--(4,0);
        \draw[fill=yellow,thick] (0,0) circle (0.7);
        \draw[fill=yellow,thick] (2,0) circle (0.7);
        \draw[fill=yellow,thick] (-2,0) circle (0.7);
        \draw[fill=yellow,thick] (3.5,0.5) rectangle (4.5,-0.5);
        \node at (-2,0) {$D_{k_3}$};
        \node at (0,0) {$C_{k_2}$};
        \node at (2,0) {$B_{k_1}$};
        \node at (4,0) {$L$};
    \end{tikzpicture}
    \label{A3-quiver}
\end{align}
Such a quiver gauge theory was not among the list of allowed theories realized with only diagonal boundary conditions, and we do expect that the general boundaries can give more quiver gauge theories in particular in the even higher-rank cases. It remains as a future work to classify all the possible quiver gauge theories from the higher-rank open spin chains with general integrable boundaries. 

\subsection{Bethe/Gauge correspondence in open XYZ model}

At the end of this section, let us discuss the potential connection between XYZ spin chain with open boundaries and 4d gauge theories with BCD-type gauge groups. 

We compare the BAE derived for the open XYZ in \cite{Yang:2005ab,Yang:2011wj,Yang:2011ff},
\begin{align}
    -\prod_{k=1}^M\frac{\sigma(u_j-u_k+\eta)\sigma(u_j+u_k+\eta)}{\sigma(u_j-u_k-\eta)\sigma(u_j+u_k-\eta)}=\frac{\sigma^{2L}(u_j+\eta/2)\sigma(2u_j+\eta)}{\sigma^{2L}(u_j-\eta/2)\sigma(2u_j-\eta)}\cr
    \times \prod_{\gamma=\pm}\prod_{l=1}^3\frac{\sigma(u_j-\alpha_l^{(\gamma)}-\eta/2)}{\sigma(u_j+\alpha_l^{(\gamma)}+\eta/2)},\label{BAE-oXYZ}
\end{align}
with six boundary parameters $\{\alpha^{(\pm)}_l\}_{l=1}^3$ against the vacua equations of 4d $\cN=1$ SO/Sp theories compactified on the torus. From the effective twisted potential \eqref{4d-potential}, the vacua equation for an SO($2k+\delta$) theory with $N_f$ fundamental matters reads 
\begin{align}
\prod_{j\neq i}^k \frac{\theta(e^{2\pi i(-\Phi_i\pm\Phi_j+m_{adj})};p)}{\theta(e^{2\pi i(\Phi_i\pm\Phi_j+m_{adj})};p)}\prod_{f=1}^{N_f}\frac{\theta(e^{-2\pi i\Phi_i+2\pi im_f};p)}{\theta(e^{2\pi i\Phi_i+2\pi im_f};p)}\frac{\theta^\delta(e^{2\pi i(-\Phi_i+m_{adj})};p)}{\theta^\delta(e^{2\pi i(\Phi_i+m_{adj})};p)}e^{E_{SO(2k+\delta)}(\Phi_i)}=1,\label{4dSO-vacua}
\end{align}
and for an Sp($k$) theory, it is given by 
\begin{align}
\prod_{j\neq i}^k \frac{\theta(e^{2\pi i(-\Phi_i\pm\Phi_j+m_{adj})};p)}{\theta(e^{2\pi i(\Phi_i\pm\Phi_j+m_{adj})};p)}\prod_{f=1}^{N_f}\frac{\theta(e^{-2\pi i\Phi_i+2\pi im_f};p)}{\theta(e^{2\pi i\Phi_i+2\pi im_f};p)}\frac{\theta(e^{2\pi i(-2\Phi_i+m_{adj})};p)}{\theta(e^{2\pi i(2\Phi_i+m_{adj})};p)}e^{E_{Sp(k)}(\Phi_i)}=1,
\end{align}
where 
\begin{align}
    E_{SO(2k+\delta)}=\frac{2\pi i(1+\varsigma)}{\varsigma}\lt(N_f+4(k-1)+\delta\rt)\Phi_i-\frac{4\pi i}{\varsigma}\lt(\sum_{f=1}^{N_f}m_f+\lt[4(k-1)+\delta\rt]m_{adj}\rt)\Phi_i, 
\end{align}
and 
\begin{align}
    E_{Sp(k)}=\frac{2\pi i(1+\varsigma)}{\varsigma}\lt(N_f+4k\rt)\Phi_i-\frac{4\pi i}{\varsigma}\lt(\sum_{f=1}^{N_f}m_f+4km_{adj}\rt)\Phi_i.
\end{align}
We shall set the mass parameters appropriately to kill the unwanted phase factor $E_{SO(2k+\delta)}$ or $E_{Sp(k)}$. On top of that, we can match the vacua equation of SO($2k+\delta$) gauge theory \eqref{4dSO-vacua} with the BAE of open XYZ spin chain \eqref{BAE-oXYZ} by choosing e.g. 
\begin{align}
    \alpha^+_1=0,\quad \alpha^+_2=\frac{1}{2},\quad \alpha^+_3=\frac{\tau}{2},\quad \alpha^-_1=\frac{1+\tau}{2},\quad \alpha^-_2=\alpha^-_3=\infty,
\end{align}
for $\delta=0$, and 
\begin{align}
    \alpha^+_1=0,\quad \alpha^+_2=\frac{1}{2},\quad \alpha^+_3=\frac{\tau}{2},\quad \alpha^-_1=\frac{1+\tau}{2},\quad \alpha^-_2=\infty,\quad \alpha^-_3=\frac{\eta}{2},
\end{align}
for $\delta=1$, where we used the elliptic version of the double-angle formula, 
\begin{equation}
    \theta_0(2z,\tau)=\theta_0(z,\tau)\theta_0\lt(z+\frac{1}{2},\tau\rt)\theta_0\lt(z+\frac{\tau}{2},\tau\rt)\theta_0\lt(z+\frac{1+\tau}{2},\tau\rt).
\end{equation}
On the other hand, similar to the reason presented in section \ref{s:off-diag}, there is still difficulty to realize the corresponding boundary contributions to 4d Sp($k$) theories from open XYZ spin chain models.

\section{Conclusion and Discussions}\label{s:conclusion}

In this article, we compared the Bethe ansatz equations of open spin chain models with the vacua equations of supersymmetric gauge theories. In our approach, each node in the quiver gauge theory is realized by an $\mathfrak{su}(2)$ layer in the nested Bethe ansatz method. We see that the basic ingredient of the quiver is SO($2k$)-type gauge nodes ($k$ is arbitrary and given by the number of magnons in the corresponding layer), as to have an SO($2k+1$)-type node, at least one boundary parameter is needed, and each Sp($k$)-type node in 2d quiver theory requires two boundary parameters to be tuned, and four parameters needed in 3d, eight parameters necessary in 4d. In XXZ $A_N$ open spin chains with diagonal boundary operators, there are only two tunable boundary parameters, and at most one Sp-type gauge node can be realized only in the 2d theory which corresponds to the XXX limit of the spin chain. There are more available boundary parameters in open spin chains with off-diagonal boundaries, whose BAEs can be worked out from the off-diagonal Bethe ansatz method developed in \cite{Cao:2013nza}, but we see that in $A_1$ open XXZ spin chain, it is still not enough to reproduce an Sp($k$) gauge theory in 3d. In the $A_N$ XXX open spin chains with general boundaries, we showed with one example that the BAEs can give rise to more different types of quiver gauge theories in 2d than those obtained from diagonal boundaries, while it is still clear that not any quiver gauge theories (even in 2d) can be reproduced from the BAEs constructed from the fusion approach of the boundaries. At the end, we also investigated the BAEs of $A_1$ XYZ open spin chains to see that it is still possible to realize the SO-type gauge theories in 4d but difficulties were again found for Sp-type gauge theories. Since our work only focused on the comparison between the BAEs of the spin chains and the vacua equations of gauge theories, there are still a lot of open problems and mysteries to be clarified in the future works. 

The Bethe/Gauge correspondence for open spin chains has so far only been investigated at the level of the equality of the equations (i.e. the Bethe ansatz equation and the vacua equation). As pointed out in \cite{Ding:2023auy,Ding:2023lsk,Ding:2023nkv}, there will be several ways to realize such Bethe/Gauge correspondences if it is just to match the equations. This instead suggests that one may use the efficient numerical methods developed in the context of spin chains, such as the rational Q-systems \cite{Marboe:2016yyn,Hou:2023ndn,Hou:2023jkr} and the Gr\"obner basis \cite{Jiang:2017phk}, to study the properties of the gauge theories reproduced in the current context near their vacua. If we are more serious about the correspondence, then it will be important to work out the wavefunction of the Bethe states in the gauge theory or to identify the dualities such as the Seiberg-like dualities of the gauge theories in the context of spin chains. 

Let us briefly explain the Seiberg-like duality in gauge theories and how it may be realized in the corresponding spin chain. The Seiberg-like dualities together with the mirror symmetry have been discussed in details in \cite{Gaiotto:2013bwa} in the context of Bethe/Gauge correspondence between A-type 3d gauge theories and closed XXZ spin chains. In the simplest case, a U($k$) gauge theory with $N_f$ fundamental and anti-fundamental flavors is dual to a U($N_f-k$) gauge theory with the same number of flavors. In the language of the spin chain, the corresponding duality is realized as a spin-flipping symmetry of Bethe states. The Seiberg-like dualities in BD-type gauge theories, however, are more complicated, as one can further consider different choices as SO($k$), O$_\pm$($k$) gauge theory with different ways to impose a $\mathbb{Z}_2$ orbifold \cite{Hori:2011pd,Aharony:2013kma,Kim:2017zis}. In 2d theories (with only $N_f$ fundamental matters), the duality is proposed to be \cite{Hori:2011pd} 
\begin{equation}
\begin{aligned}
    {\rm O}_+(k)\leftrightarrow {\rm SO}(N_f-k+1),\\
    {\rm SO}(k)\leftrightarrow {\rm O}_+(N_f-k+1),\\
    {\rm O}_-(k)\leftrightarrow {\rm O}_-(N_f-k+1).
\end{aligned}\label{Seiberg-d}
\end{equation}
Let us explain the above duality in BAE-like equations with one simple example, O$_+$($2$) theory with $5$ fundamental flavors dual to SO($4$) theory with the same number of matters, based on the argument presented in \cite{Hori:2011pd}. The vacua equations (i.e. BAE-like equations) are given by 
\begin{equation}
    (-1)^{N_f+1}\prod_{i=1}^{N_f}\frac{\sigma_a-\tilde{m}_i}{-\sigma_a-\tilde{m}_i}=1,\quad a=1,\dots,k.\label{BAE-noadj}
\end{equation}
For each $\sigma_a$, there are $N_f$ solutions to the equation, but we need to further impose some identification under the Weyl group including the permutation among $\sigma_a$'s and the flip, 
\begin{equation}
    \sigma_a\to \epsilon_a\sigma_a,\quad \epsilon_a=\pm 1.
\end{equation}
When $N_f$ is odd, one of the solutions to each of the equation \eqref{BAE-noadj} is $0$, and the remaining solutions are paired in the form $(\mu_i,-\mu_i)$ ($i=1,\dots,\frac{N_f-1}{2}$). There are two sectors of vacua: one does not contain $\sigma_a=0$ and the other has one of $\sigma_a$'s vanishing. For O($k$) groups, all the flip identifications $\sigma_a\to -\sigma_a$ are independently imposed, and in the first sector there are $\lt(\begin{array}{c}
\frac{N_f-1}{2}\\
\frac{k}{2}
\end{array}\rt)$ vacua, and the second sector has $\lt(\begin{array}{c}
\frac{N_f-1}{2}\\
\frac{k}{2}-1
\end{array}\rt)$ ground states. For SO($k$) groups with even $k$, only the identifications satisfying $\prod_{a=1}^{k/2}\epsilon_a=1$ are imposed, and there are $2\lt(\begin{array}{c}
\frac{N_f-1}{2}\\
\frac{k}{2}
\end{array}\rt)$ vacua in the first sector with all $\sigma_a$ non-zero, $\lt(\begin{array}{c}
\frac{N_f-1}{2}\\
\frac{k}{2}-1
\end{array}\rt)$ in the second sector. An ambiguity in the O($k$) theory is that there are two ways to implement the $\mathbb{Z}_2$ orbifold in chiral multiplet with twisted mass. One is to simply identify $\mathbb{Z}_2:\Phi\to-\Phi$, and the other further includes a twisting with respect to the fermion number $(-1)^{F_s}$, i.e to identify the action $\mathbb{Z}_2(-1)^{F_s}$. This choice in particular affects the second sector of vacua with one of $\sigma_a=0$, and in the theory denoted as O$_+(k)$ with the standard orbifold choice, both twisted and untwisted vacua survive with the number of vacua doubled to $2\lt(\begin{array}{c}
\frac{N_f-1}{2}\\
\frac{k}{2}-1
\end{array}\rt)$. In the non-standard theory denoted as O$_-(k)$, only the twisted one survives from the projection. In total, for $N_f$ odd, O$_+(k)$ theory has $\lt(\begin{array}{c}
\frac{N_f-1}{2}\\
\frac{k}{2}
\end{array}\rt)+2\lt(\begin{array}{c}
\frac{N_f-1}{2}\\
\frac{k}{2}-1
\end{array}\rt)$ vacua, O$_-(k)$ theory has $\lt(\begin{array}{c}
\frac{N_f-1}{2}\\
\frac{k}{2}
\end{array}\rt)+\lt(\begin{array}{c}
\frac{N_f-1}{2}\\
\frac{k}{2}-1
\end{array}\rt)$ vacua, and SO($k$) theory has $2\lt(\begin{array}{c}
\frac{N_f-1}{2}\\
\frac{k}{2}
\end{array}\rt)+\lt(\begin{array}{c}
\frac{N_f-1}{2}\\
\frac{k}{2}-1
\end{array}\rt)$. In this way, we see that the number of vacua matches in O$_+(2)$ theory with $N_f=5$ ($2+2\times 1=4$ vacua) and SO($4$) theory with $N_f=5$ ($2\times 1+2=4$ vacua), and the Seiberg-like duality can then be established between pairs of ground states. In the current context with gauge theories dual to spin chains, an adjoint chiral multiplet is further contained in the gauge theory, and the Seiberg-like duality \eqref{Seiberg-d} needs to be modified accordingly as lower-dimensional versions of \cite{Leigh:1995qp}. We plan to investigate such dualities more thoroughly from the viewpoint of the symmetries of $Q$-functions in open spin chains as a future work.

Unlike the correspondence between closed spin chains and A-type gauge theories, open spin chains have free parameters from the boundary operators that do not have obvious corresponding on the gauge theory side. It thus strongly suggests that the Bethe/Gauge correspondence for BCD-type gauge theories and open spin chains is very likely not a one-to-one correspondence. It is then more tempting to turn to the open integrable model associated to Maulik-Okounkov's R-matrix (and Intermediate Long Wave hydrodynamics) built in \cite{Litvinov:2021phc}, in which it is claimed that the boundary operator $K$ might only have three solutions that corresponds to BCD-type ${\cal W}$-algebras respectively and the obtained integrable systems describe a BCD-version of the Bazhanov-Lukyanov-Zamolodchikov integrable structure \cite{Bazhanov:1994ft} of affine Toda QFT. Similarly the connection has been explored in \cite{Bonelli:2015kpa} between the integrable systems (of spin Intermediate Long Wave hydrodynamics) and gauge theories on ALE spaces of A- and D-types. Tighter constraints may only appear in the generic $\Omega$-background with an enhanced quantum algebraic symmetry, and it is desirable to work out the details of the relation between the open integrable systems and Nekrasov's instanton partition functions of BCD-type gauge theories, potentially with the help of the Alday-Gaiotto-Tachikawa-Wyllard relation \cite{Alday:2009aq,Wyllard:2009hg} that connects supersymmetric gauge theories with gauge group $G$ and 2d CFTs with $G$-type ${\cal W}$-algebraic symmetry \cite{Keller:2011ek}.

\paragraph{Acknowledgement}

We would like to thank Yunfeng Jiang, Taro Kimura, Hongfei Shu, Wenli Yang, Peng Zhao, Kilar Zhang and Hao Zou for inspiring discussions on relevant topics. 
R.Z. is supported by National Natural Science Foundation of China No. 12105198 and the High-level personnel project of Jiangsu Province (JSSCBS20210709).

\appendix

\section{Effective twisted potential from 4d $\cN=1$ Theory on $D^2_\epsilon\times T^2$}\label{a:4d-part}

In this Appendix, we first summarize the results of \cite{Longhi:2019hdh} on the 4d partition function on $D^2_\epsilon\times T^2$ and then derive the effective 4d twisted potential from the partition function. The partition function is given as a contour integral over the Jeffrey-Kirwan poles, 
\ba
Z=\oint{\rm d}\Phi Z_{BPS}(\Phi)=\oint{\rm d}\Phi Z^{vec}_{1-loop}(\Phi)Z^{chiral}_{1-loop}(\Phi),
\ea
where more explicitly we have 
\ba
Z^{vec}_{1-loop}(\Phi)=\lt(\frac{e^{-\frac{\pi i}{3}P_3(0)}}{\Res_{u=0}\Gamma(u,p,q^2)}\rt)^{{\rm rank}(G)}\det_{adj}{}^\prime\lt(\frac{e^{-\frac{\pi i}{3}P_3(\Phi)}}{\Gamma(\Phi,p,q^2)}\rt),
\ea
as the one-loop determinant for the vector multiplet, and for the chiral multiplet with the so-called Robin-like boundary condition, the one-loop determinant is given by, 
\ba
Z^{chiral,R}_{1-loop}(\Phi)=\det_R\lt(e^{\frac{\pi i}{3}P_3(\tau r/2+\Phi)}\Gamma(\frac{\tau r}{2}+\Phi,p,q^2)\rt).
\ea
In correspondence to the 3d case, we adopted the notation $q:=e^{\pi i\tau}=e^{-\beta_2}$ ($\tau=\frac{i}{\pi}\beta_2$), and $p:=e^{2\pi i\varsigma}$ characterizing the modulus of $T^2$. $r$ is the R-charge of the scalar in the chiral multiplet. We also used the double $q$-factorial, 
\ba
(x;p,q)_\infty:=\prod_{i,j\geq 0}(1-p^i q^jx),
\ea
to define the elliptic $\Gamma$-function 
\ba
\Gamma(z;p,q):=\frac{(pqe^{-2\pi iz};p,q)_\infty}{(e^{2\pi iz};p,q)_\infty}.
\ea
The polynomial $P_3$ is defined as 
\ba
P_3(x)=B_{33}(x|1,\tau,\varsigma)-\frac{1-\tau^2+\tau^4}{24\varsigma(\tau+\tau^2)},
\ea
with 
\ba
B_{33}(x|1,\tau,\varsigma)&:=&\frac{x^3}{\tau\varsigma}-\frac{3(1+\tau+\varsigma)x^2}{2\tau\varsigma}\nn\\
&&+\frac{1+\tau^2\varsigma^2+3(\tau+\varsigma+\tau\varsigma)}{2\tau\varsigma}x-\frac{(1+\tau+\varsigma)(\tau+\varsigma+\tau\varsigma)}{4\tau \varsigma}.
\ea

To implement the $\Omega$-background deformation with a parameter $\epsilon$, one can simply replace $\beta_2\rightarrow \beta_2/\epsilon$ in the partition function as discussed in the remark of section 3.1 in \cite{Kimura:2020bed}. The effective twisted potential can be worked out from the asymptotic behavior of the partition function $Z$ in the limit $\epsilon\rightarrow 0$, 
\begin{equation}
    Z\sim \exp\lt(\frac{1}{\epsilon}W^{eff}(\Phi)\rt).
\end{equation}
Note that we have the following identity, 
\begin{equation}
    (x;p,q)_\infty=e^{-Li_3(x;p,q)},\quad Li_3(x;p,q)=\sum_{k=1}^\infty \frac{x^k}{k(1-p^k)(1-q^k)},
\end{equation}
and then by using 
\ba
\log \det A={\rm tr}\log A,
\ea
we obtain in the classical limit $\epsilon\rightarrow 0$ (with $q=e^{-\beta_2\epsilon}$ rescaled) that 
\begin{align}
    \epsilon \log\Gamma(x,p,q^2)=\epsilon\lt(Li_3(e^{2\pi ix};p,q^2)-Li_3(pq e^{-2\pi ix};p,q^2)\rt)\cr
    \rightarrow \frac{1}{2\beta_2}\sum_{k=1}^\infty \frac{e^{2\pi ik x}}{k^2(1-p^k)}-\frac{1}{2\beta_2}\sum_{k=1}^\infty \frac{p^ke^{-2\pi ik x}}{k^2(1-p^k)}
=:g(2\pi ix;p).
\end{align}
It is then straightforward to derive the effective twisted potential is as 
\ba
W^{eff}(\Phi)=-\sum_\alpha g(2\pi i\alpha\cdot\Phi;p)-\frac{\pi^2}{3\beta_2\varsigma}\sum_{\alpha}(\alpha\cdot\Phi)^3+\frac{\pi^2(1+\varsigma)}{2\beta_2\varsigma}\sum_\alpha(\alpha\cdot \Phi)^2\nn\\
+\sum_w g(2\pi iw\cdot\Phi)+\frac{\pi^2}{3\beta_2\varsigma}\sum_{w}(w\cdot\Phi)^3-\frac{\pi^2(1+\varsigma)}{2\beta_2\varsigma}\sum_w(w\cdot \Phi)^2,
\ea
where $\alpha$ runs over all the roots in the gauge group, $w$ stands for the weights appearing in the representation of the chiral multiplet, and we used
\begin{align}
    \epsilon \frac{\pi i}{3}P_3(x)\rightarrow -\frac{\pi^2}{12\beta_2}(1+\varsigma)-\frac{\pi^2}{72\beta_2\varsigma}+\frac{\pi^2}{3\beta_2\varsigma}x^3-\frac{\pi^2(1+\varsigma)}{2\beta_2\varsigma}x^2+\frac{\pi^2(1+3\varsigma)}{6\beta_2\varsigma}x,
\end{align}
with $\tau\rightarrow \tau \epsilon$ to include the effect of $\Omega$-background. One may also turn on the flavor symmetry of the chiral multiplets by replacing $w\cdot \Phi\rightarrow w\cdot \Phi+m_f$.

\bibliographystyle{JHEP}
\bibliography{open}

\end{document}